# Hot Carrier Trapping Induced Negative Photoconductance in InAs Nanowires towards Novel Nonvolatile Memory

Yiming Yang, [1§] Xingyue Peng,[1] Hong-Seok Kim, [2§] Taeho Kim,[2] Sanghun Jeon,[2] Hang Kyu Kang,[3] Wonjun Choi,[3] Jindong Song,[3] Yong-Joo Doh,[2*] and Dong Yu[1*]

[1]Department of Physics, University of California, Davis, California 95616, United States

[2]Department of Applied Physics, Korea University Sejong Campus, Sejong 339-700, Republic of Korea

[3]Center for Opto-Electronic Materials and Devices Research, Korea Institute of Science and Technology, Seoul 136-791, Republic of Korea

§ These authors contributed equally to this work.

*E-mail: yjdoh@korea.ac.kr, yu@physics.ucdavis.edu

**Keywords:** Nanowire, hot carriers, negative photoconductance, indium arsenide, scanning photocurrent microscopy, nonvolatile memory

**We report a novel negative photoconductivity (NPC) mechanism in n-type indium arsenide nanowires (NWs). Photoexcitation significantly suppresses the conductivity with a gain up to $10^5$. The origin of NPC is attributed to the depletion of conduction channels by light assisted hot electron trapping, supported by gate voltage threshold shift and wavelength dependent photoconductance measurements. Scanning photocurrent microscopy excludes the possibility that NPC originates from the NW/metal contacts and reveals a competing positive photoconductivity. The conductivity recovery after illumination substantially slows down at low temperature, indicating a thermally activated detrapping mechanism. At 78 K, the spontaneous recovery of the conductance is completely quenched, resulting in a reversible memory device which can be switched by light and gate voltage pulses. The novel NPC based optoelectronics may find exciting applications in photodetection and nonvolatile memory with low power consumption.**

Indium arsenide (InAs), a semiconductor with a direct bandgap of 0.35 eV, has found many applications including infrared photodetectors, photovoltaics, and diode lasers. Thanks to their high carrier mobilities, InAs nanowires (NWs) have recently attracted much attention as a model system for studying low dimensional quantum transport[1,2] and as a material for high-speed optoelectronic applications.[3-6] On the other hand, InAs is known to have a surface electron accumulation layer[7] that can have major impacts on charge transport, manifested as random telegraph noise and large field effect hysteresis.[8,9] Thus, understanding the nature of the surface states is therefore critical to both fundamental studies and novel device applications. Most existing work on investigating the surface effects of InAs NWs mainly uses electronic measurements, while the combination of optical and electronic techniques may provide additional valuable insights that cannot be gathered from pure electronic measurements. In this work, we apply spatially and spectrally resolved optoelectronic techniques to understand the surface states of InAs NWs. Specifically, we focus on exploring the negative photoconductance (NPC) behavior of InAs NWs and use it to probe the spatial and energetic distribution of the surface states.

In most semiconductors, photoexcitation injects additional charge carriers which lead to an increase in conductivity. However, in uncommon cases, conductivity may decrease under photoexcitation, known as NPC. NPC has been observed in a variety of semiconductor nanostructures, such as p-type carbon nanotubes,[10] Bi doped p-type ZnSe nanowires,[11] n-type InN thin films,[12] etc. In the cases of p-type systems, NPC has been attributed to the photon assisted oxygen desorption from the nanostructure surface, which results in a decrease of hole concentration and hence conductivity. This mechanism has been experimentally supported by the observation of a long conductivity recovery time in vacuum in seconds or minutes after light is turned off. However, such an explanation does not apply to the NPC observed in n-type systems, as oxygen desorption would only lead to higher electron concentration and conductivity. Indeed, n-type ZnO NWs, also showing a slow recovery in vacuum, only exhibits positive photoconductivity (PPC).[13] Nevertheless, NPC has also been observed in n-type semiconductors such as in n-type InN thin films.[12] Its origin has been attributed to enhanced carrier scattering by light-induced positively charged gap states, evidenced by a reduced mobility in the presence of light.

Here, we report a study of the NPC in field effect transistors (FETs) incorporating single n-type InAs NWs. The origin of the NPC is fundamentally different from other systems reported previously. The NPC observed in n-type InAs NWs is unlikely caused by $O_2$ desorption as in the p-type systems; and the mobilities of InAs NWs appear to be constant upon illumination, different from the InN thin films. Instead, we attribute the NPC in InAs NWs to photoexcitation induced majority carrier (electron) trapping, which leads to a reduction in free carrier concentration. More surprisingly, band-edge excitation only creates PPC and only photons with energy much higher than the bandgap can induce NPC. In addition, at low temperature, the conductance does not recover after illumination but can be turned on with a gate voltage pulse. Taking advantage of the reversible bistable states, we further demonstrate NPC based nonvolatile memories.

InAs NW arrays were synthesized by ultrahigh vacuum molecular beam epitaxy (Figure 1a, see *Methods* for more details). The NWs were single crystalline with growth axis along the (111) direction, confirmed by transmission electron microscopy (TEM) (Figure 1d). A typical single InAs NW FET device (Figure 1e) demonstrates linear current-voltage ($I$-$V_{sd}$) curves both in dark and under illumination (Figure 1b). The conductance decreases upon illumination, with an apparent reduction at intensity as weak as 1 mW/cm$^2$. We have measured over 30 individual devices, and all of the devices demonstrated such an NPC behavior upon illumination. We note that PPC in InAs NW devices has also been reported previously.[6, 14] However, the reported PPC seems to be caused by extrinsic electric field either at Si/InAs p-n junctions[6] or at the possible Schottky contacts to highly resistive InAs NWs.[14] The NPC observed in our InAs NWs is intrinsic, not influenced by the contact Schottky field as demonstrated by spatially resolved photocurrent imaging (which will be shown later). Photoexcitation in the Si bottom gate may lead to electron accumulation at the defect states in Si or the interface of Si, which can reduce the conductance of the InAs NW channel. However, we excluded this possibility, since the NW devices made of other materials with similar mobility and bandgap, such as PbS, do not show NPC.[15] We did not observe a correlation of NPC with carrier mobility in InAs NWs. High mobility InAs NW devices up to $1.2 \times 10^4$ cm$^2$/(V s) also exhibit NPC behaviors (Figure S1 in Supporting Information). The photoconductance ($\Delta G = G - G_0 < 0$, where $G$ is the conductance under illumination and $G_0$ is the conductance in the dark) is linear at low intensity but saturates as higher intensity (Figure 1f).

Quantitatively, at low intensity the responsivity is estimated to be $R = \Delta I / P = -3 \times 10^4$ A/W at $V_{sd} = 0.2$ V and the incident wavelength $\lambda = 500$ nm, where $\Delta I = I - I_0$ is the photocurrent and $P$ is the incident power, estimated by the light projected upon the physical cross section of the NW. Note that the negative sign represents the current reduction. A gain of $\Gamma = (\Delta I/e) / (P/h\nu) = R h\nu /e = -7.5 \times 10^4$ is also estimated, where $h\nu$ is the photon energy and $e$ is the magnitude of the electron charge. Both the responsivity and gain values are high, compared to devices based on PPC, including both NW and 2D nanostructure photoconductors.[14, 16-20] We summarize the gain values and photoresponse times of reported nanostructured photodetectors in Table 1.

In order to investigate the NPC mechanism, we performed gate voltage ($V_g$) sweeps under dark and light. The hysteresis of the gate sweep is negligible at the scan rate we used (1 V/s), so we only show one branch for clarity. The dark gate sweep of the InAs NW device shown in Figure 1c demonstrates n-type behavior with an electron mobility of $\mu_n = 650$ cm$^2$/(V s) and an electron concentration of $n = 2.9 \times 10^{17}$ cm$^{-3}$ at $V_g = 0$ V. The NW can be completely depleted at negative $V_g$. In the presence of illumination, the gate sweep curves shift to the right (Figure 1c). The slope of the gate sweep in the linear region stays almost unchanged at all intensities, indicating that carrier mobility remains the same under excitation. This is different from the previous reported NPC in InN thin films,[12] where mobility is reduced under illumination. Instead, the NPC in our case is induced by a shift of $V_g$ threshold ($V_{th}$). To be more quantitative, we extract $V_{th}$ by extrapolating the gate sweep curve as shown in Figure 1c. $V_{th}$ increases rapidly at low intensity and much more slowly at higher intensity (Figure 1g).

Next we performed scanning photocurrent microscopy (SPCM) measurements to obtain the spatial distribution of NPC. A reasonably fast photoresponse is required for meaningful SPCM studies. Among the measured devices, the response time varies significantly in a range of sub-milliseconds to seconds at room temperature. The conductance turn-off upon the onset of illumination is always faster than the conductance turn-on after the illumination is turned off. To test the response time in the fast devices, an optical chopper is introduced in the light path, and the current signal is measured by a digital oscilloscope. When illumination is turned on, the conductance drops faster than 0.1 ms (limited by the temporal resolution of our preamplifier). When the illumination is turned off, the conductance also recovers quickly in < 0.1 ms but followed by a slower tail with a characteristic time constant of about 5 ms (Figure 2a). The photoresponse in a slower device is shown in Figure S2 in Supporting Information. We choose fast-response devices for performing SPCM measurements.

At $V_{sd}$ = 0V, SPCM shows photocurrent spots of different polarities near the contacts (Figure S3 in Supporting Information), caused by band bending at contacts.[21-23] In this *letter*, to study NPC, we focus on the photoconductance at non-zero $V_{sd}$. Figure 2b demonstrates a typical SPCM image for a fast device with $V_{sd}$ = 10 mV. The SPCM image clearly confirms that NPC originates from the InAs NW body rather than the NW/metal contacts, as the entire NW has nearly uniform NPC. The photocurrent dip in the vertical cross section manifests NPC (Figure 2d). The dip has a width much larger than the width of the laser spot (~300 nm), reflecting the photoresponse is not fast enough to follow the scanning beam (beam moves from top to bottom in about 0.1 s). The slower side of the asymmetric dip on the right can be attributed to the slower conductance recovery process, in which the laser has moved away from the wire but the conductance is not fully recovered yet. Interestingly, at higher intensity, an additional narrower central peak emerges, in addition to the broader NPC dip (Figure 2c, e). This indicates that the photoconductance has a PPC component in addition to the NPC component. The total photoconductance is a sum of the two components, *i.e.*, $\Delta G = \Delta G_P + \Delta G_N$. PPC has a faster photoresponse than NPC. $\Delta G_P$ is much smaller than $|\Delta G_N|$ at low intensity so that SPCM image shows a simple single dip. At high intensity, $\Delta G_P$ is comparable to $|\Delta G_N|$ so that a more complicated "W"-shaped photocurrent cross section is observed (Figure 2e). As the laser spot moves from top to bottom (Figure 2c), the NW is first illuminated by the low intensity tail of the Gaussian beam, where the PPC component can be ignored and the NPC dominates (point A). When the center of the laser spot moves onto the NW, the PPC component becomes comparable to the NPC in magnitude, leading to an increase of photocurrent (point B). Finally, as the laser spot moves down away from the NW, the fast PPC component decays quickly and the NPC dominates again (point C). The different response times of the two components thus allow us to extract both components from the SPCM images. As the NPC remains constant at the bottom of the dip with a relative large width (Figure 2d), we can estimate $\Delta G_N$ from the maximum decrease and $\Delta G_P$ from the central peak height as indicated in Figure 2e. We then measured both components from SPCM images taken at different excitation intensity (Figure 2f). The PPC component increases almost linearly with intensity, while the magnitude of the NPC component increases rapidly at low intensity but saturates at about 3 W/cm$^2$ (Figure 2f inset).

We further explore the wavelength dependence of NPC. A global illumination (intensity is uniform along the NW) at low intensity is used so that PPC component is small and can be ignored. Figure 3a shows the time trace of photoconductance as the light with the same intensity but different wavelength is turned on and off. As seen, the magnitude of the photoconductance substantially decreases at longer wavelength. Figure 3b summarizes the magnitude of the photoconductance as a function of incident wavelength. Interestingly, though InAs has a bandgap of 0.35 eV, the NPC completely disappears when the wavelength is longer than 1300 nm (~ 1 eV) (Figure 3b). SPCM images taken at different wavelengths (Figure 3c) further confirmed such observations. At short wavelength ($\lambda$ = 532 nm) both NPC and PPC components are observed in the SPCM image, while at longer wavelength ($\lambda$ = 1500 nm) only a PPC component is observed. The NPC at 532 nm is manifested by the background color contrast, which is not seen at 1500 nm. On the other hand, the central peaks in both figures indicate the existence of PPC at both wavelengths. The wavelength threshold for observing NPC varies slightly in the range of 900 - 1300 nm from device to device but the disappearance of NPC at long incident wavelength is observed in all devices.

Now we summarize the main experimental observations: (i) The gate voltage threshold shifts to more positive values under illumination, while the carrier mobility remains nearly unchanged. (ii) The total photoconductivity is composed of two competing components: a fast PPC component and a slower NPC component. NPC quickly saturates with light intensity, while PPC is proportional to intensity. (iii) Only photons with energy much higher than the bandgap can create NPC. Then we discuss the possible mechanism for the observed NPC in n-type InAs NWs: observation (i) indicates that the NPC mechanism in our n-type InAs NWs is different from that in InN thin films, in which case the mobility significantly changes under illumination.[12] The gate threshold change is most likely caused by the electrostatic screening by light induced charge trapping at the surface of InAs NWs or at the interface of InAs and $SiO_2$. The distribution of the charge traps is relatively uniform along the NW axis as indicated by the uniform NPC observed in SPCM images. The direction of the gate threshold shift under illumination signals electron trapping, which lowers the electron density in the n-type conduction channel. At higher intensity, the gate threshold changes much more slowly (Figure 1g), indicating that traps are filled up by charges. This is consistent with the saturation of NPC at high intensity in observation (ii) (Figures 1f and 2f). We estimate the trap density to be on the order of $10^{12}$ cm$^{-2}$ by $C_g \Delta V_{th} / e A$, where $C_g$ is the gate dielectric capacitance, $A$ is the surface area of the NW (assuming the charge traps are uniformly distributed at the NW surface), and $\Delta V_{th}$ is the maximum change of the gate threshold by light and is estimated to be 4 V from Figure 1g. This is about 0.1% of the atom surface density and comparable to the previous report.[24] For observation (ii), the fast PPC component is consistent with fast band-edge recombination of the photo-induced carriers. NPC has a slower photoresponse, particularly with a slow conductance recovery tail (Figure 2a), indicating that the trapped charges are slowly released to the conduction band. The response time varies from device to device, indicating a device (and surface) dependent detrapping time. Finally, observation (iii) suggests only hot electrons can be trapped to induce NPC. The mean free path of a hot electron in high-mobility InAs NWs is expected to be comparable to the NW diameter, allowing time for hot electrons to reach the surface traps before thermalization occurs. Electrons excited by low energy photons are likely blocked by an energy barrier from the trap states and hence only PPC is observed. Above the photon energy threshold, the photogenerated hot electrons can overcome the barrier and get trapped. After illumination, the trapped electrons can be released back to the conduction band over the energy barrier likely through a thermal activation process. Note that the photoconductance spectrum below the wavelength threshold (Figure 3b) does not necessarily follow a thermal activation behavior (exponential with photon energy). This is because the photoconductance also depends on the wavelength-dependent photogeneration rate and the trapping process of the hot electrons may not be purely thermal activation, different from the detrapping process.

The above discussion is summarized into an energy diagram shown in Figure 3d. Briefly, a high energy photon generates hot electrons in InAs, which may be trapped before thermally relaxing to the band edge. Trapped electrons electrostatically reduce the electron density in the conduction band and lower conductance. After the light is turned off, the trapped electrons may slowly return to the conduction band and the photo-suppressed conductance recovers. The surface of InAs has been known to have a thin layer of $In_2O_3$ ($E_g \sim 3$ eV). It is possible that the photogenerated electrons are trapped in the oxide layer. Further investigation is needed in order to determine the exact chemical nature of the traps. The threshold of ~1 eV for NPC corresponds to a trap barrier height of ~0.6 eV (as it needs to subtract the InAs bandgap and the kinetic energy of a hot hole). The spontaneous recovery of conductance at room temperature imply that the detrapping barrier height is a few times $k_B T \sim 0.026$ eV, where $k_B$ is the Boltzmann constant and $T$ is the room temperature. Therefore, the energy level of the trap states is most likely ~0.5 eV above the conduction band.

To further understand the NPC mechanism, we propose a model involving hot electron trapping. The characteristic time scales for trapping and detrapping are denoted as $\tau_{trp}$ and $\tau_{dtrp}$. The thermalization and band edge recombination time scales are $\tau_{th}$ and $\tau_r$. We assume an equal $\tau_{th}$ for electrons and holes for simplicity. Thus we establish a set of equations to describe the dynamics of involved physical processes, as depicted in Figure 3d:

$$\frac{dn_{hot}}{dt} = G - \frac{n_{hot}}{\tau_{th}} - \frac{n_{hot}}{\tau_{trp}}\left(1 - \frac{n_{trp}}{N_{trp}}\right) \qquad (1)$$

$$\frac{dn_{trp}}{dt} = \frac{n_{hot}}{\tau_{trp}}\left(1 - \frac{n_{trp}}{N_{trp}}\right) - \frac{n_{trp}}{\tau_{dtrp}} \tag{2}$$

$$\frac{d\Delta n}{dt} = \frac{n_{hot}}{\tau_{th}} - \frac{\Delta p}{\tau_r} + \frac{n_{trp}}{\tau_{dtrp}} \tag{3}$$

$$\frac{dp_{hot}}{dt} = G - \frac{p_{hot}}{\tau_{th}} \tag{4}$$

$$\frac{d\Delta p}{dt} = \frac{p_{hot}}{\tau_{th}} - \frac{\Delta p}{\tau_r} \tag{5}$$

These differential equations describe the change rates of the densities of photogenerated hot electrons, trapped electrons, band-edge electrons, hot holes, and band-edge holes, respectively. For example, equation (1) means that the change rate of photogenerated hot electrons ($\frac{dn_{hot}}{dt}$) is given by the competition between generation rate ($G$), electron thermalization ($\frac{n_{hot}}{\tau_{th}}$), and trapping $\frac{n_{hot}}{\tau_{trp}}\left(1 - \frac{n_{trp}}{N_{trp}}\right)$. The definitions of variables in the above equations are listed in Table 2. Note that the recombination rate is determined by minority carriers ($\Delta p/\tau_r$) in equation (3). The steady state equations then do not directly contain $\Delta n$ but $\Delta n$ can be found through the charge neutrality equation ($\Delta n + n_{hot} + n_{trp} = \Delta p + p_{hot}$), which is implied by equations (1)-(5) (see derivation in Supporting Information).

As a rough estimation, we assume the mobility of hot carriers is the same as band-edge carriers so the photoconductivity can be expressed as

$$\Delta\sigma = (\Delta n + n_{hot})e\mu_n + (\Delta p + p_{hot})e\mu_p \tag{6}$$

We assume steady state and consider two extreme cases:

**Case I**: low intensity photoexcitation such that most of the trap states are unoccupied and available. In this case, we have $n_{trp} \ll N_{trp}$ and (derivation can be seen in Supporting Information)

$$\Delta\sigma = Ge(\mu_n + \mu_p)(\tau_{th} + \tau_r) - Ge\mu_n \frac{\tau_{th}\tau_{dtrp}}{\tau_{th}+\tau_{trp}} \quad (7)$$

Usually $\tau_{th}$ is the shortest time scale in the above expression and in n-type semiconductors electron conduction dominates ($\mu_n \gg \mu_p$). So the photoconductivity is reduced to

$$\Delta\sigma = Ge\mu_n \left(\tau_r - \frac{\tau_{th}\tau_{dtrp}}{\tau_{trp}}\right) \quad (8)$$

The first term accounts for the normal positive photoconductivity, while the second term accounts for the negative photoconductivity caused by electron trapping. Intuitively, the photoexcitation results in free electrons (majority carriers) being trapped, which lowers free electron concentration and conductivity. In other word, photoexcitation effectively "moves" free electrons to localized trap states and therefore reduces the conductivity. As shown in equation (8), the magnitude of the negative photoconductivity term increases with the electron detrapping time $\tau_{dtrp}$, since electrons stay trapped for a longer time at larger $\tau_{dtrp}$; and decreases with the electron trapping time $\tau_{trp}$, since electrons more likely thermalize to the band edge rather than get trapped at larger $\tau_{trp}$. Correspondingly, the gain has an expression

$$\Gamma = \frac{\tau_r - \frac{\tau_{th}\tau_{dtrp}}{\tau_{trp}}}{\tau_{tr}} = \frac{\tau_r - \tau'}{\tau_{tr}} \quad (9)$$

where $\tau_{tr}$ is the transit time across the channel length and $\tau' = \tau_{th}\tau_{dtrp}/\tau_{trp}$. In addition to the normal PPC gain ($\tau_r/\tau_{tr}$), the additional term $-\tau'/\tau_{tr}$ in equation (9) originates from electron trapping, which leads to a reduced free electron concentration in the conduction band as required by charge neutrality. Indeed, from equations (1)-(5), we find (derivation can be seen in Supporting Information):

$$\Delta n = G(\tau_r - \frac{\tau_{th}\tau_{dtrp}}{\tau_{trp}}) = G(\tau_r - \tau') \qquad (10)$$

which suggests $\Delta n < 0$, if $\tau' > \tau_r$. This confirms the free electron density ($n = n_0 + \Delta n$, where $n_0$ is dark density and $\Delta n < 0$ is the photoinduced change) is reduced by photoexcitation. The gain is negative, as long as $\tau' > \tau_r$. Quantitatively, with a channel length of $l = 4$ μm and electron mobility of $\mu_n = 650$ cm$^2$/(V s) extracted from the gate dependence measurements, we estimate the transit time to be $\tau_{tr} = 2.5$ ns. $\tau_r$ for InAs is expected to be on the order of 1-10 ns and hence the PPC component has a gain on the order of 1, much smaller than the magnitude of the observed NPC. The large gain magnitude of the NPC indicates $\tau' >> \tau_r$ such that $\Gamma = (\tau_r - \tau')/\tau_{tr} \approx -\tau'/\tau_{tr}$. From the measured gain value, we have estimated $\tau' = 0.2$ ms, indeed much longer than $\tau_r$. If assuming a thermalization time of $\tau_{th} \sim 1$ ps, $\tau_{dtrp}/\tau_{trp} = \tau'/\tau_{th} \sim 10^9$. This large ratio of trapping and de-trapping rates is expected because of the energy barrier blocking the trapped electrons from returning to the conduction band.

**Case II**: high intensity photoexcitation such that nearly all trap states are occupied, $n_{trp} \approx N_{trp}$. In this case, (see derivation in Supporting Information)

$$\Delta\sigma = Ge(\mu_n + \mu_p)(\tau_{th} + \tau_r) - N_{trp}e\mu_n \qquad (11)$$

where the first term is the PPC component and the second term is the NPC component, which is saturated to a constant value $-N_{trp}e\mu_n$, in agreement with the experimental results.

In order to confirm the existence of an energy barrier and further understand the detrapping mechanism, we performed photoconductance measurements at low temperature. At 78 K, as the illumination is turned on, the conductance is immediately decreased. After the illumination is turned off, the conductance does not spontaneously recover (Figure 4a), and the device remains in the high resistive state (HRS) in the dark for a long time (>1000 s). The conductance ratio of the HRS and the low resistive state (LRS) is up to 5 orders of magnitude (Figure 4b). Gate sweep indicates that the gate threshold is shifted to positive voltage after illumination similar to the room temperature results (Figure 4c). Interestingly, the conductance can be recovered by scanning gate voltage to a negative voltage (Figure 4d). These results are consistent with our proposed mechanism involving an energy barrier between the trap states and the conduction channel. The detrapping process is likely a thermally activated process and is greatly quenched at low temperature. The application of a negative gate voltage, on the other hand, may assist the releasing of the trapped carriers by electrostatic repulsion. This observation also further excludes the possible origination of NPC by light-induced oxygen desorption: a gate voltage can be used to recover the conductance, which cannot be explained by oxygen adsorption.

The large conductivity on/off ratio, the stability of bistable states, and the rewritability by application of a gate voltage suggest that the NPC behavior in InAs NWs may be applied to novel nonvolatile memory devices. Compared with the Ovshinsky type of nonvolatile devices based on order-disorder phase transition,[25-27] one advantage of InAs NW devices is the low power consumption since the NPC is very sensitive to light with a gain of about $10^5$. As shown in our model, the high gain value in the NPC devices is intrinsically associated with a long $\tau'$ = $\tau_{th}\tau_{dtrp}/\tau_{trp}$. For application, reproducible and fast switching is required. We demonstrate that the device can be turned on and off by over 50 cycles without any sign of fatigue (Figure 4e). For fast switching, one may use a short gate voltage pulse (~0.1 s) instead of gate scanning to recover the device to LRS (Figure 4f), in a similar fashion as removing persistent photoconductivity in a metal oxide photo-sensor.[28] Furthermore, the conductance of LRS can be controlled by the magnitude of the gate pulse (Figure 4f), enabling multistate memory switching. Though we demonstrated the memory behavior at 78 K, the recovery of the photo-suppressed conductance at 250 K takes hundreds of seconds (Figure S4 in Supporting Information), already much slower than at room temperature. Surface engineering to increase the barrier height may suppress the thermal detrapping process and enable operation at room temperature.

In summary, we have observed sensitive negative photoconductivity in n-type high mobility InAs NW FETs. The spatially and spectrally resolved photocurrent studies reveal that the conductance suppression is caused by light induced hot electron trapping. The recovery of the conductance after illumination is a thermally assisted process, fast at room temperature but completely quenched at low temperature. The device remains almost permanently insulating at 78 K after illumination but the conductance can be recovered by applying a negative gate voltage pulse. This work provides insights on understanding charge trapping in InAs NWs and may also lead to novel optical memory devices with low power consumption.

**Methods**

Undoped InAs NWs were grown on a (111)-oriented Si substrate by ultrahigh vacuum molecular beam epitaxy in the presence of gold particles (~ 50 nm diameters) as a catalyst. The Si substrate with gold nanoparticles was heated up to 850 $^{\circ}$C to remove surface oxide in the main chamber under $10^{-9}$ Torr pressure. After reflection high-energy electron diffraction (RHEED) observation of 2 folded pattern, the substrate temperature was decreased to around 450 °C. It was found that the density of gold nano-particles was preserved during the deoxidation of Si substrate. The growth of InAs was started by injecting arsenic tetramer for 5 minutes, followed by simultaneous introduction of In and arsenic tetramers. The fluxes of arsenic tetramer were at $2\times10^{-6}$ Torr. The fluxes of In were equivalent to the growth rates of InAs (0.09 nm/s). In this method, the length of the free-standing NWs was obtained approximately 5 μm with diameters varying from 40 to 50 nm. Although there are defects including rotational twins and mixtures of zinc-blende and wurtzite structures around 1 μm length region from the gold nanoparticles, the other parts of NWs to the 4 μm length region from the bottom show uniform crystal structures of wurtzite with a small number of rotational twins as shown in Figure 1d, which was taken around the 2 μm points from the gold nanoparticles.

To fabricate single NW FETs, the as-grown InAs NWs were transferred to a highly p-doped silicon substrate covered by a 300 nm-thick oxide layer. The conductive silicon substrate was used as a back gate electrode. Source and drain electrodes were formed by conventional electron-beam lithography and subsequent electron-beam evaporation of a Ti/Au (10/120 nm) bilayer. Prior to the metal deposition, the NWs were treated in a highly dilute (0.5 %) ammonium polysulfide, $(NH_4)_2S_x$, solution for ohmic contact.[29]

The detailed description of our scanning photocurrent microscopic (SPCM) setup can be found in our previous reports.[15, 22, 30] Briefly, a laser beam is focused to a diffraction limited spot and is raster scanned on the NW device substrate, while the photocurrent is recorded as a function of laser position and plotted as a two dimensional (2D) image. The wavelength dependence measurements are performed using a super-continuum laser (NKT) with an acousto-optic tunable filter (AOTF). The output wavelength can be continuously tuned in the range of 500-900 nm and 1100-1700 nm with a spectral width of 1 nm. The repetition rate of the tunable laser is 80 MHz with a pulse width of 5 ps. The high repetition rate makes the tunable laser pseudo-CW. We also examined the NPC effect with a CW laser (Coherent Compass 315M-100) at $\lambda$ = 532 nm. Under the same photon energy and illumination intensity, we did not observe any difference when using the pulsed laser or the CW laser.

**Supporting Information**

NPC behaviors for a high mobility device, slower photoresponse, SPCM image at $V_{sd} = 0$ V, photoresponse at 250 K, and derivation in the carrier trapping model. The Supporting Information is available free of charge on the ACS Publications website at http://pubs.acs.org.

**Notes**

The authors declare no competing financial interest.

**Acknowledgements**

This work was supported by the U.S. National Science Foundation Grant DMR-1310678, the Basic Science Research Program (NRF Grant 2012R1A2A1A01008027) and the KIST Flag Ship and Future Convergence Pioneer Program. Work at the Molecular Foundry was supported by the Office of Science, Office of Basic Energy Sciences, of the U.S. Department of Energy under Contract No. DE-AC02-05CH11231. We thank J. Y. Song for useful discussions.

**Table 1**. Photoconductive gain, response time and operation spectrum of various 1D and 2D nanostructure photodetectors.

| photodetectors | type | gain | response time | spectrum | reference |
|---|---|---|---|---|---|
| InAs NW | NPC | $10^5$ at 0.2 V | < 5 ms | visible | this work |
| InAs NW | PPC | $10^3$ at 3 V | < 10 ms | near IR | 14 |
| Ge NW | PPC | $10^3$ at 4 V | < 0.1 ms | visible | 18 |
| ZnO NW | PPC | $10^8$ at 5 V | > 10 s | UV | 19 |
| InP NW | PPC | $10^4$ at 0.05V | < 3 ms | visible | 17 |
| 2D MoS$_2$ | PPC | 0.02 at 1 V | 50 ms | visible | 20 |
| Graphene | PPC | $10^{-3}$ at 4 mV | ~ ps | near IR | 16 |

**Table 2.** Physical parameters used in the hot electron trapping model.

| | |
|---|---|
| $G$ | photoexcitation rate |
| $n_{hot}, p_{hot}$ | density of hot electrons/holes |
| $\Delta n, \Delta p$ | density of electrons/holes at the band edge |
| $n_{trp}$ | density of trapped electrons |
| $N_{trp}$ | density of trap states |
| $\tau_{th}$ | hot carrier thermalization time |
| $\tau_r$ | band-edge carrier recombination lifetime |
| $\tau_{trp}$ | trapping time |
| $\tau_{dtrp}$ | detrapping time |

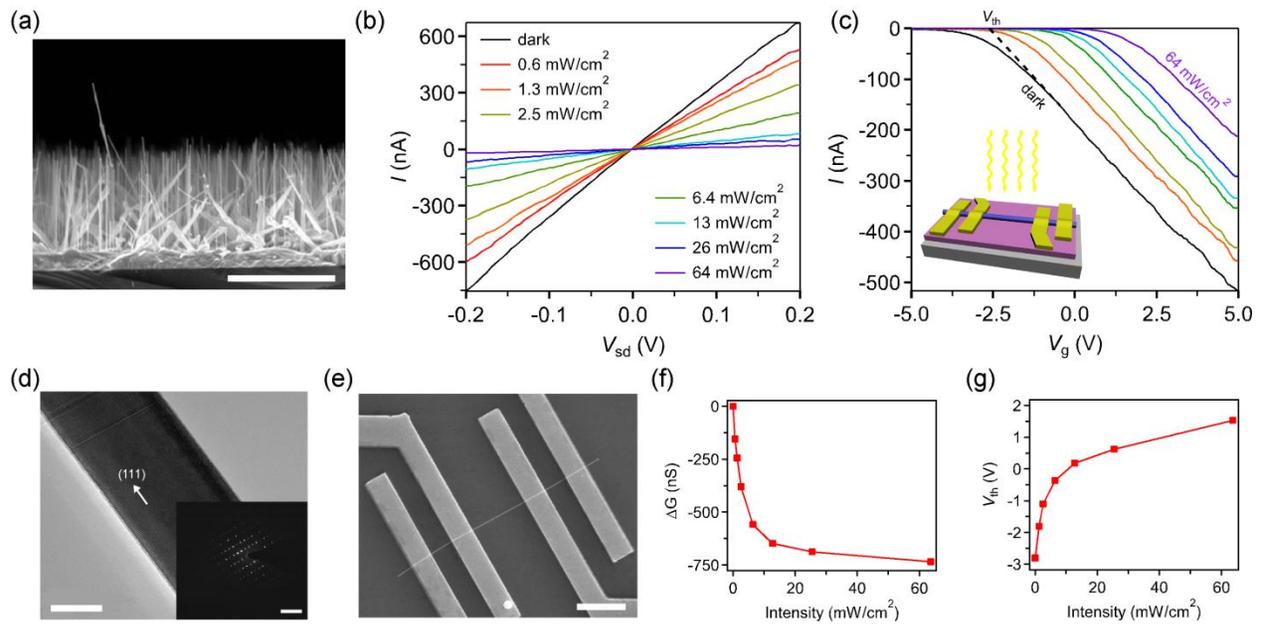

**Figure 1**. Charge transport characteristics and NPC behaviors of InAs NW FETs. (a) Scanning electron microscope (SEM) image of the vertical cross section of an as-grown InAs NW array. Scale bar: 5 μm. (b) $I$-$V_{sd}$ curves in the dark and under various illumination intensity. The wavelength of the light is 500 nm. All the $I$-$V_{sd}$ curves were measured at $V_g$ = 0 V. (c) Gate sweep of the NW FET under dark and light at $V_{sd}$ = 50 mV. The light intensity of each curve is shown in the legends of (b). The dashed line demonstrates the extrapolation of the gate threshold $V_{th}$. Inset: schematic drawing of the device setup. (d) TEM image of a single InAs NW. Scale bar: 20 nm. Inset: selected area electron diffraction (SAED) patterns showing wurtzite structure over the entire region. Scale bar: 2 nm$^{-1}$. (e) SEM image of a four-probe InAs single NW FET. Scale bar: 2 μm. (f) Conductance plotted as a function of light intensity extracted from (b). (g) Gate threshold $V_{th}$ versus intensity as extracted from (c).

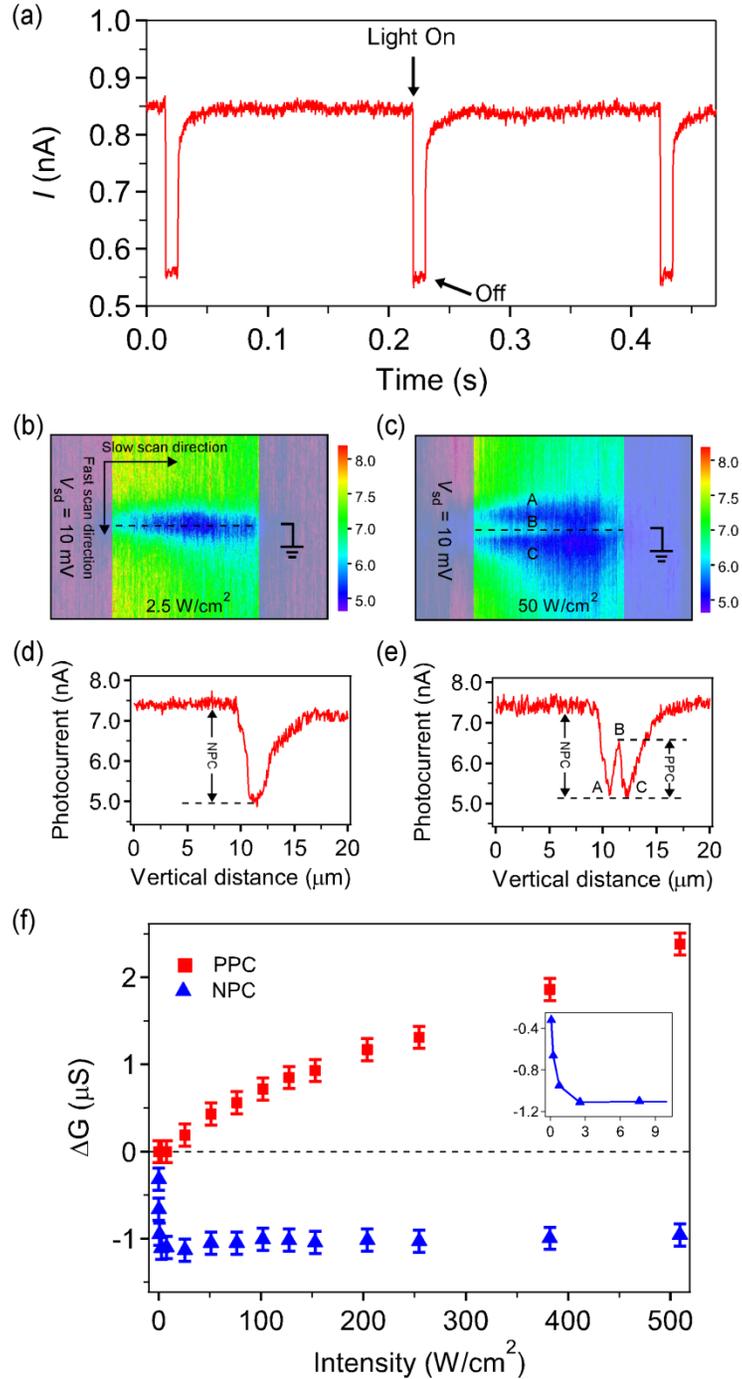

**Figure 2**. Intensity dependence of photoconductance. (a) Time trace of conduction current as the light is turned on and off. (b, c) SPCM images at $V_{sd} = 10$ mV with a laser spot peak intensity at 2.5 and 50 W/cm$^2$, respectively. The unit of current is in nA. (d, e) Vertical cross section of the SPCM image in (b) and (c). (f) NPC and PPC components as a function of intensity. Inset: zoom-in of (f) in the low intensity range.

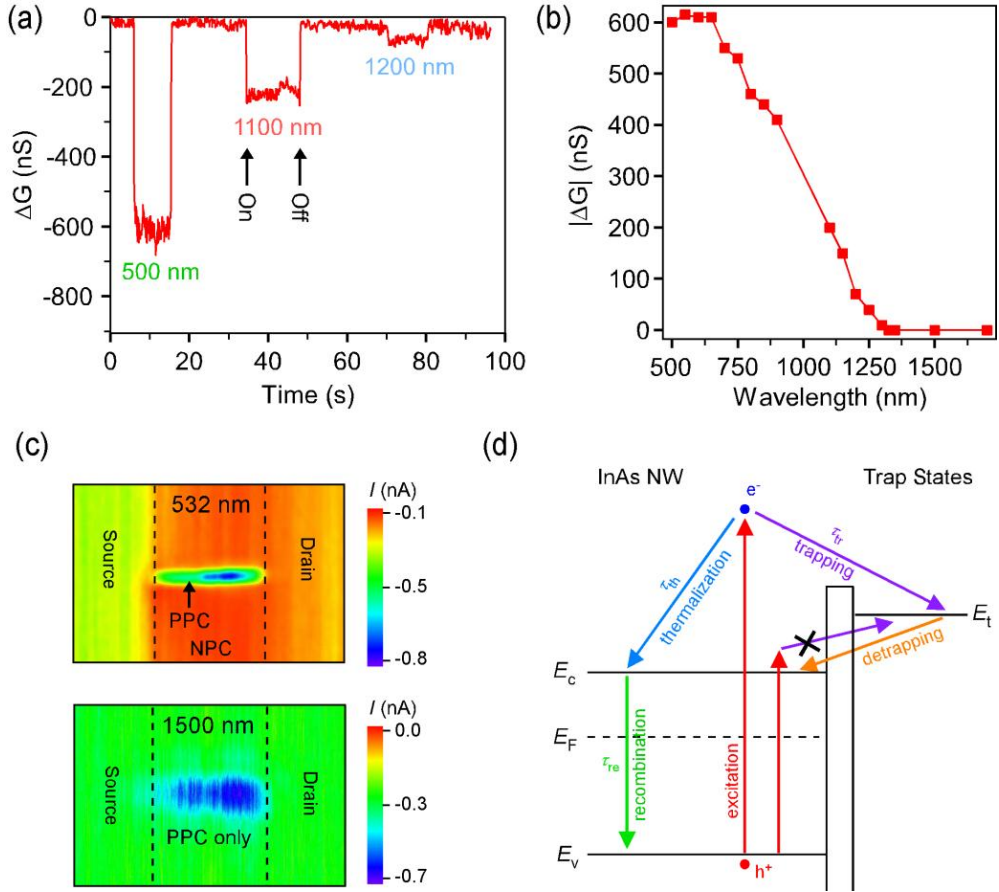

**Figure 3**. Wavelength dependence of photoconductance. (a) Time trace of photoconductance as light with the same intensity (~0.1 W/cm$^2$) but at different wavelength is turned on and off. (b) Magnitude of the photoconductance as a function of wavelength. (c) SPCM images under 532 and 1500 nm illumination at $V_{sd}$ = -50 mV. The source electrode is connected to $V_{sd}$ and the drain electrode is connected to preamp. NPC is manifested in the top figure by the background color contrast between the electrode areas and the central area. This color contrast is missing in the bottom figure. PPC is observed in both figures. (d) Band diagram showing excitation, thermalization, recombination, trapping and detrapping processes. $E_c$, $E_v$, $E_F$, and $E_t$ represent conduction band, valence band, Fermi level, and trap level, respectively. An energy barrier separates the trap states from the conduction channel.

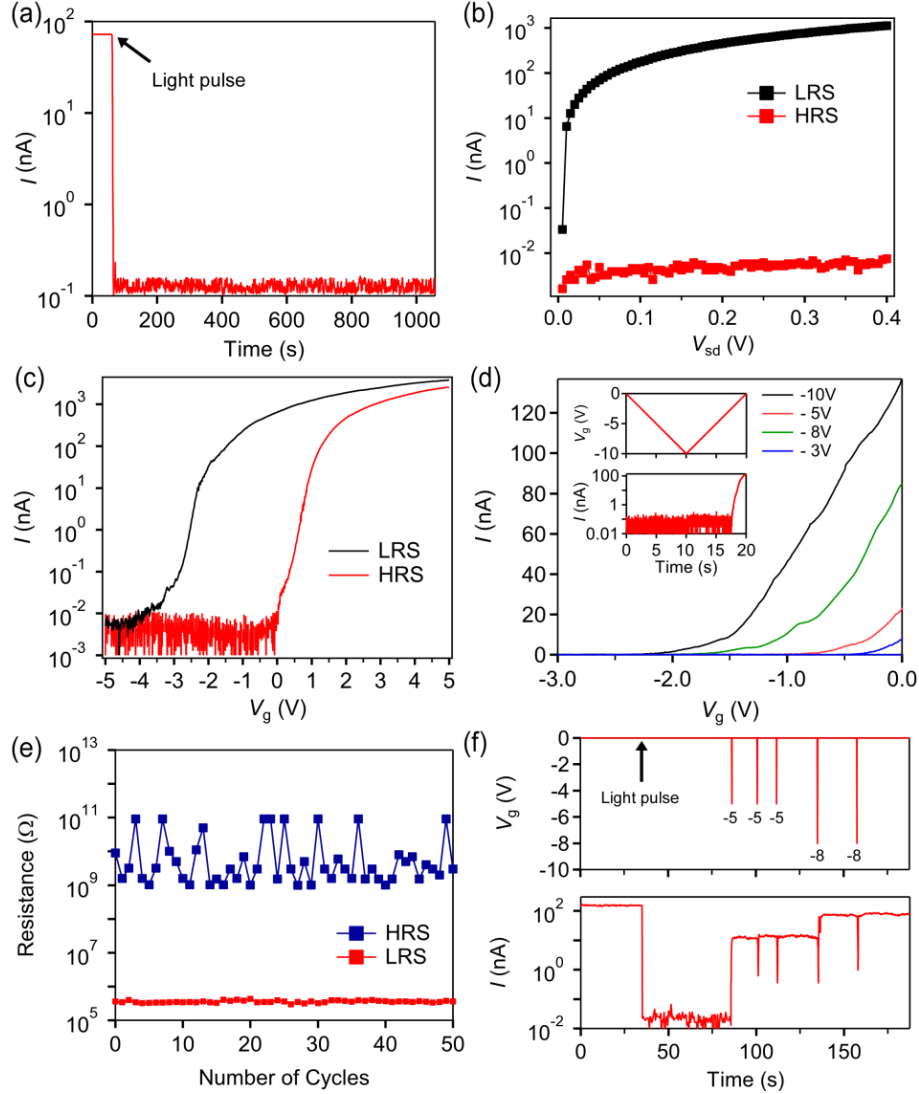

**Figure 4.** Optical memory effect at low temperature (all data in this figure is taken at 78 K). (a) Time trace of current at $V_{sd}$ = 0.1 V. After a light pulse, the insulating device does not regain conductivity. (b) $I$-$V_{sd}$ curves before (LRS) and after (HRS) light pulse. (c) Gate sweep before and after light pulse at $V_{sd}$ = 0.1 V. $V_g$ scans from 5 V to -5 V. (d) Conductance recovery by scanning $V_g$ to a negative value. Different curves are for different minimum $V_g$. For example, the black curve is for a gate scan from 0 V to -10 V and then back to 0 V as indicated by the inset. The conductance is turned on as the gate scans back to 0 V. (e) Memory fatigue test: the device is turned off by light and then recovered by applying a gate voltage pulse. This process is repeated by 50 cycles. (f) Gate pulse with a width of about 0.1 s is used to turn on the device. Pulse with different height ($V_g$ = -5 V or - 8V) can switch the device to different conductive states.